\documentclass[5p,sort&compress]{elsarticle}
\usepackage{amssymb, amsmath,color, hyperref, graphicx}
\usepackage[switch]{lineno}
\begin{document}
\newcommand {\todo}[1] {\textcolor{red}{#1}}

\begin{frontmatter}

\title{Excited bottomonia in quark-gluon plasma from lattice QCD}


\author[1]{Rasmus Larsen \corref{cor1}}
\ead{rlarsen@bnl.gov}

\author[2,3]{Stefan Meinel }
\ead{smeinel@email.arizona.edu}

\author[1]{Swagato Mukherjee}
\ead{swagato@bnl.gov}

\author[1]{Peter Petreczky}
\ead{petreczk@quark.phy.bnl.gov}

\address[1]{Physics Department, Brookhaven National Laboratory, Upton, NY 11973, USA}
\address[2]{Department of Physics, University of Arizona, Tucson, Arizona 85721, USA}
\address[3]{RIKEN-BNL Research Center, Brookhaven National Laboratory, Upton, NY 11973, USA}

\cortext[cor1]{Corresponding author}

\begin{abstract}

We present the first lattice QCD study of up to $3S$ and $2P$ bottomonia at non-zero
temperatures. Correlation functions of bottomonia were computed using novel
bottomonium operators and a variational technique,  within the lattice non-relativistic
QCD framework. We analyzed the bottomonium correlation functions based on simple
physically-motivated spectral functions. We found evidence of sequential in-medium
modifications, in accordance with the sizes of the bottomonium states.

\end{abstract}
\end{frontmatter}
\section{Introduction}\label{sc:intro}

Quarkonium suppression has been proposed as a signature of quark-gluon plasma (QGP)
formation in heavy-ion collisions~\cite{Matsui:1986dk}. The main idea behind this
proposal was the observation that color screening within a deconfined medium can make
the interaction between the heavy quark and anti-quark short ranged, leading to the
dissolution of quarkonia in QGP. At a given temperature, different quarkonium states
are expected to be affected differently by QGP--- a more tightly bound quarkonia
having a smaller size is less influenced by the medium than a relatively loosely
bound, larger one. Therefore, following the hierarchy of their binding energy and
sizes, a sequential pattern of in-medium modification is
expected~\cite{Karsch:1987pv,Digal:2001ue}. Some evidence of sequential in-medium
modification of quarkonia comes from lattice QCD studies of S-wave and P-wave
quarkonium correlators along the
temporal~\cite{Aarts:2010ek,Aarts:2013kaa,Aarts:2014cda,Kim:2018yhk,Kim:2014iga} and
spatial~\cite{Bazavov:2014cta,Karsch:2012na} directions. Recent studies have revealed
that inclusion of dissipative effects lead to  a more complex theoretical picture of
in-medium heavy quark and anti-quark
interactions~\cite{Laine:2006ns,Brambilla:2008cx}. However, the main conclusion,
i.e., that quarkonium dissolve in QGP when the temperature is large enough compared to its
inverse size and binding energy,  have remained
unchanged~\cite{Laine:2007gj,Burnier:2007qm,Beraudo:2007ky,Petreczky:2010tk,Burnier:2015tda}.
Hints for sequential in-medium modification of bottomonia have also been observed in
heavy-ion collision
experiments~\cite{Chatrchyan:2011pe,Chatrchyan:2012lxa,Khachatryan:2016xxp,Sirunyan:2017lzi,Wang:2019vau}.
While the connection between the observed hierarchy of the $\Upsilon(nS)$ yields in
heavy-ion collisions and the expected sequential melting of these states in QGP is
complicated by dynamical effects, such a link is expected to exist
\cite{Krouppa:2017jlg,Yao:2017fuc,Petreczky:2016etz}. For this reason, the study of
sequential in-medium quarkonium modifications in heavy-ion collisions is a subject of
extensive experimental and theoretical efforts; for recent reviews see
Refs.~\cite{Mocsy:2013syh,Aarts:2016hap}.

There have been many attempts to study in-medium properties of
charmonium~\cite{Umeda:2002vr,Datta:2002ck,Karsch:2002wv,Datta:2003ww,Asakawa:2003re,Jakovac:2006sf,Ohno:2011zc,Ding:2012sp,Ding:2017std}
and bottomonium
\cite{Jakovac:2006sf,Aarts:2010ek,Aarts:2011sm,Aarts:2012ka,Aarts:2013kaa,Aarts:2014cda,Kim:2014iga,Kim:2018yhk}
in lattice QCD, almost entirely focused on in-medium modifications of ground states
of S- and P-wave quarkonium. Previous lattice QCD studies of in-medium quarkonium
used point meson operators, i.e., operators with quark and anti-quark fields located
in the same spatial point, which are known to have non-optimal overlap with the
quarkonium wave-functions, especially, with the excited states. As a result, these
correlators are largely dominated by the vacuum continuum parts of the spectral
function, and isolating the contributions of in-medium bottomonium becomes quite
difficult~\cite{Mocsy:2007yj,Petreczky:2010tk,Burnier:2015tda}. Recently, we explored
the possibility of studying in-medium bottomonium properties using correlators of
extended meson operators~\cite{Larsen:2019bwy}. We found that such operators have
very good overlap with the lowest S- and P-state bottomonia, thereby allowing us to
cleanly isolate the vacuum continuum contributions to the bottomonium correlators. We
showed that these correlators are more sensitive to the in-medium bottomonium
properties than the ones with point sources. Analyzing these correlators, we found evidence for
thermal broadening of the $1S$- and $1P$-state bottomonia, however, excited
bottomonium states remained elusive. In this letter we introduce novel extended meson
operators within the lattice non-relativistic QCD (NRQCD) formalism, which, for the
first time, allow us to probe in-medium modifications of up to $3S$ and $2P$
bottomonium.

\section{Methodology}\label{sc:method}

\begin{figure*}[!t]
  \centering
  \includegraphics[width=0.29\textwidth]{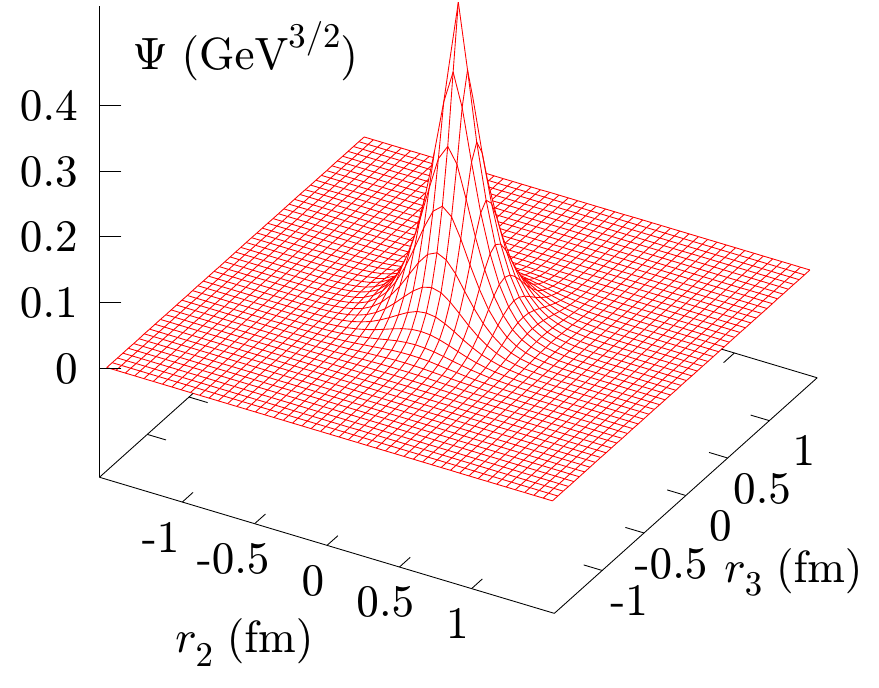}
  \hspace{0.03\textwidth}
  \includegraphics[width=0.29\textwidth]{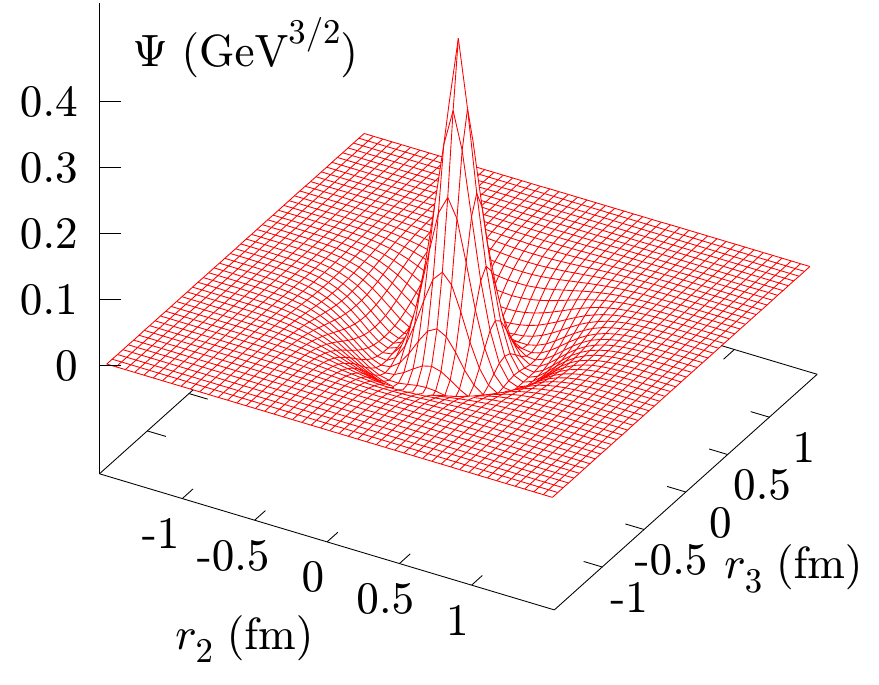}
  \hspace{0.03\textwidth}
   \includegraphics[width=0.29\textwidth]{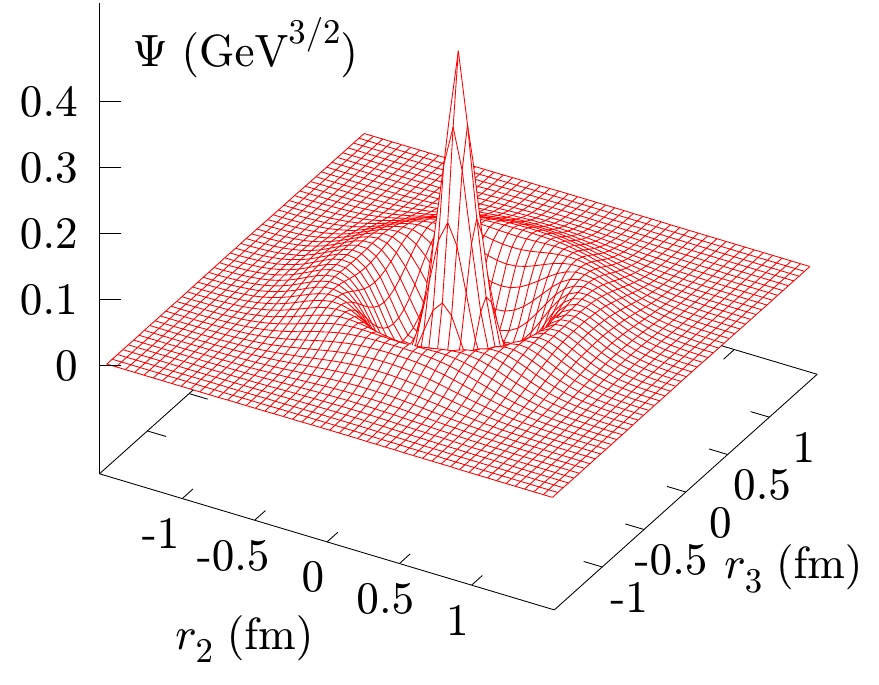}
   \caption{The shape-functions $\Psi_{1S}$ (left),  $\Psi_{2S}$ (middle),
   $\Psi_{3S}$ (right) used to calculate $\eta_b$ and $\Upsilon$ correlators for
   lattices with $a=0.0655$~fm.}
   \label{fig:phi}
\end{figure*}

The lattice NRQCD Lagrangian employed in this study is exactly the same as in
Refs.~~\cite{Larsen:2019bwy,Meinel:2010pv}--- tree-level tadpole improved, accurate
up to order $v^4$, but also includes $v^6$ spin dependent terms. For the background
gauge fields we used $2+1$-flavor gauge configurations on $48^3 \times 12$ lattices
with bare gauge couplings $\beta=6.74$, $6.88$, $7.03$, $7.28$ and $7.596$,
corresponding lattice spacings $a=0.1088$, $0.0951$, $0.0825$, $0.0655$, $0.0493$~fm and temperatures $T=151$,
$173$, $199$, $251$ and $334$~MeV, respectively. For each gauge coupling we also
carried out the corresponding vacuum $T=0$ calculations. All gauge configurations
were generated by the HotQCD collaboration~\cite{Bazavov:2011nk,Bazavov:2014pvz},
with the physical value of the strange quark mass, and up/down quark masses
corresponding to the pion mass of $160$~MeV in the continuum limit. The lattice
spacings, $a$, were determined using the $r_1$ scale from the static quark anti-quark
potential and the value $r_1=0.3106(18)$~fm~\cite{Bazavov:2010hj}. The mass parameter
in the NRQCD Lagrangian was fixed through the kinetic mass of the $\eta_b$ meson, described
in detail in Ref.~\cite{Larsen:2019bwy}.

To calculate bottomonium correlators we used novel extended meson operators in Coulomb gauge of the form
\begin{equation}
 O_i(\mathbf{x},t) = \sum_{\mathbf{r}} \Psi_i(\mathbf{r}) \bar{q}(\mathbf{x}+\mathbf{r},t)
 \Gamma q(\mathbf{x},t) \,.
\label{operator}
\end{equation}
The different choices of $\Gamma$ used in this work can be found in Table 1 of
Ref.~\cite{Larsen:2019bwy}. The index $i$ refers to the different states in a given
channel, e.g., 1S, 2S, 3S etc . The shape-functions $\Psi_i$ were obtained by solving
the discretized Schrodinger equation with a Cornell potential on a 3-dimensional
lattice having a lattice spacing and a volume exactly the same as that of the
corresponding QCD background. Spin interactions were neglected. For the lattice
Schrodinger equation we used an $O(a^2)$-improved Laplace operator.  For the Cornell
potential the string tension was chosen to be $(468~\mathrm{MeV})^2$,
and the Coulomb part was computed at tree-level in lattice perturbation theory for
the Symanzik-improved lattice gluon action with a fixed strong coupling constant
$\alpha_S = 0.24$. The bottom-quark mass was set to $m_b = 4.676$~GeV.   More details
can be found in the Appendix D of Ref.~\cite{Meinel:2010pv}. The $\Psi$'s
used for calculations of $\eta_b$ and $\Upsilon$ correlators for lattices with
$a=0.0655$~fm are shown in Fig. \ref{fig:phi}.

Since $\Psi_i$ is a good approximation for the wave-function of the $i^\mathrm{th}$
vacuum bottomonium,  as expected, the corresponding operator $O_i$ was found to have
a good overlap with the $i^\mathrm{th}$ state. However, the off-diagonal correlators,
$G_{ij}(t)=\langle O_i(t) O_j^\dagger (0) \rangle$ for $i\ne j$, were found to be non-zero,
though small. Thus, we resorted to the variational analysis by considering linear
combinations $\tilde O_{\alpha} = \Omega_{\alpha j} O_j$ such that $\langle \tilde
O_{\alpha}(t) \tilde O_{\beta}^\dagger(0) \rangle \propto \delta_{\alpha,\beta}$. The matrices
$\Omega_{\alpha j}$ were obtained using the generalized eigenvalue
problem~\cite{Nochi:2016wqg,Michael:1985ne,Luscher:1990ck,Blossier:2009kd,Orginos:2015tha}
$G_{ij}(t) \Omega_{\alpha j}=\lambda_{\alpha}(t,t_0) G_{ij}(t_0) \Omega_{\alpha j}$.
We calculated the correlators of optimized operators $C_{\alpha}(t)=\langle \tilde
O_{\alpha}(t) \tilde O_{\alpha}^\dagger(0) \rangle$ for 1S, 1P, 2S, 2P and 3S. When calculating $\Omega_{\alpha j}$, the value of
$t$ was chosen such that it corresponds to physical extent $\tau=t\cdot
a\simeq0.5$~fm and $t_0$ was chosen to be $\tau =0$ fm. Choosing $t_0 /a$ to be 1 or 2 does not change the results significantly.

\section{Results}\label{sc:results}

\subsection{Vacuum case}\label{sc:vacuum}

\begin{figure*}[!t]
  \centering
  \includegraphics[width=0.43\textwidth]{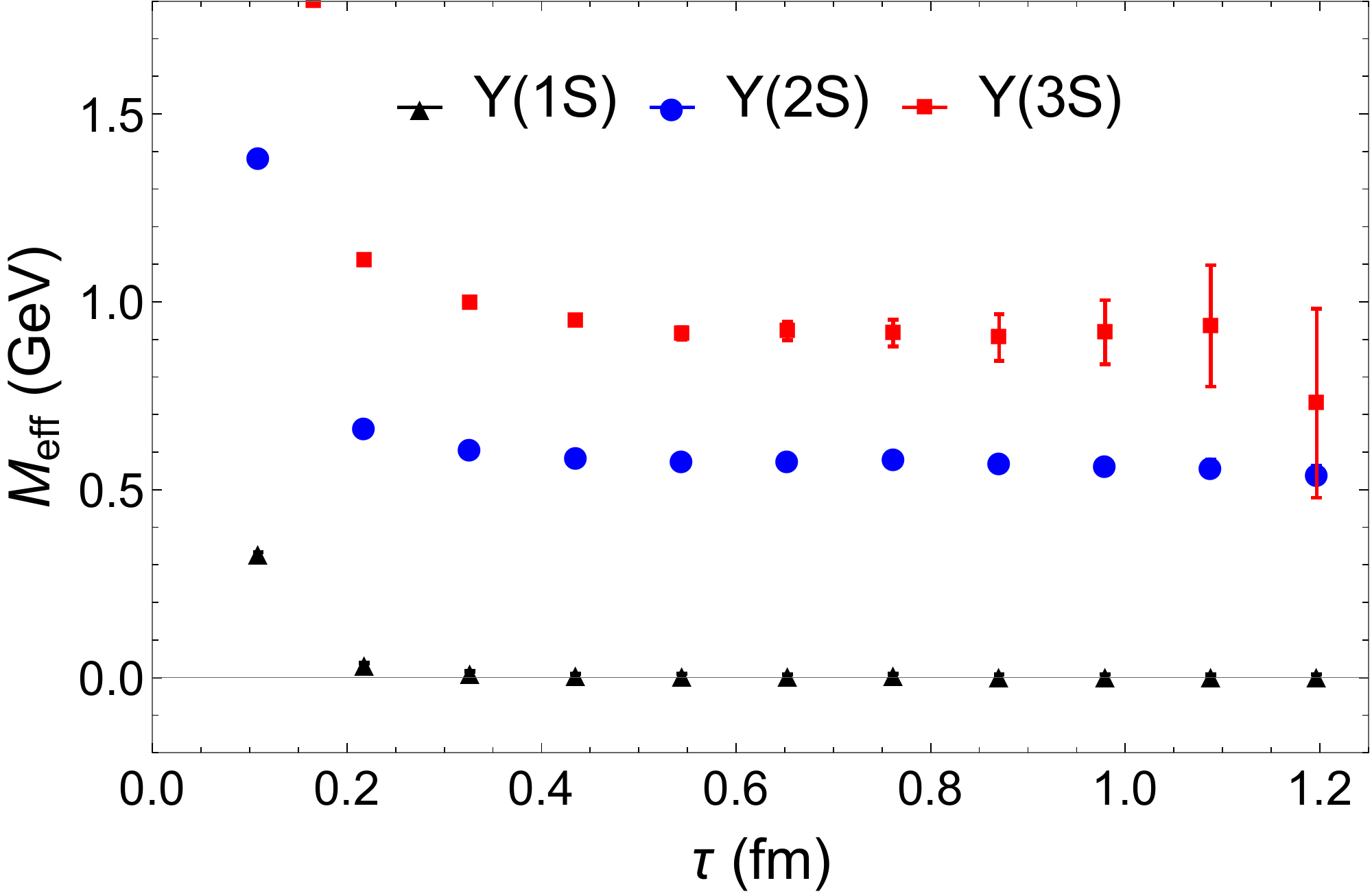}
  \hspace{0.03\textwidth}
  \includegraphics[width=0.43\textwidth]{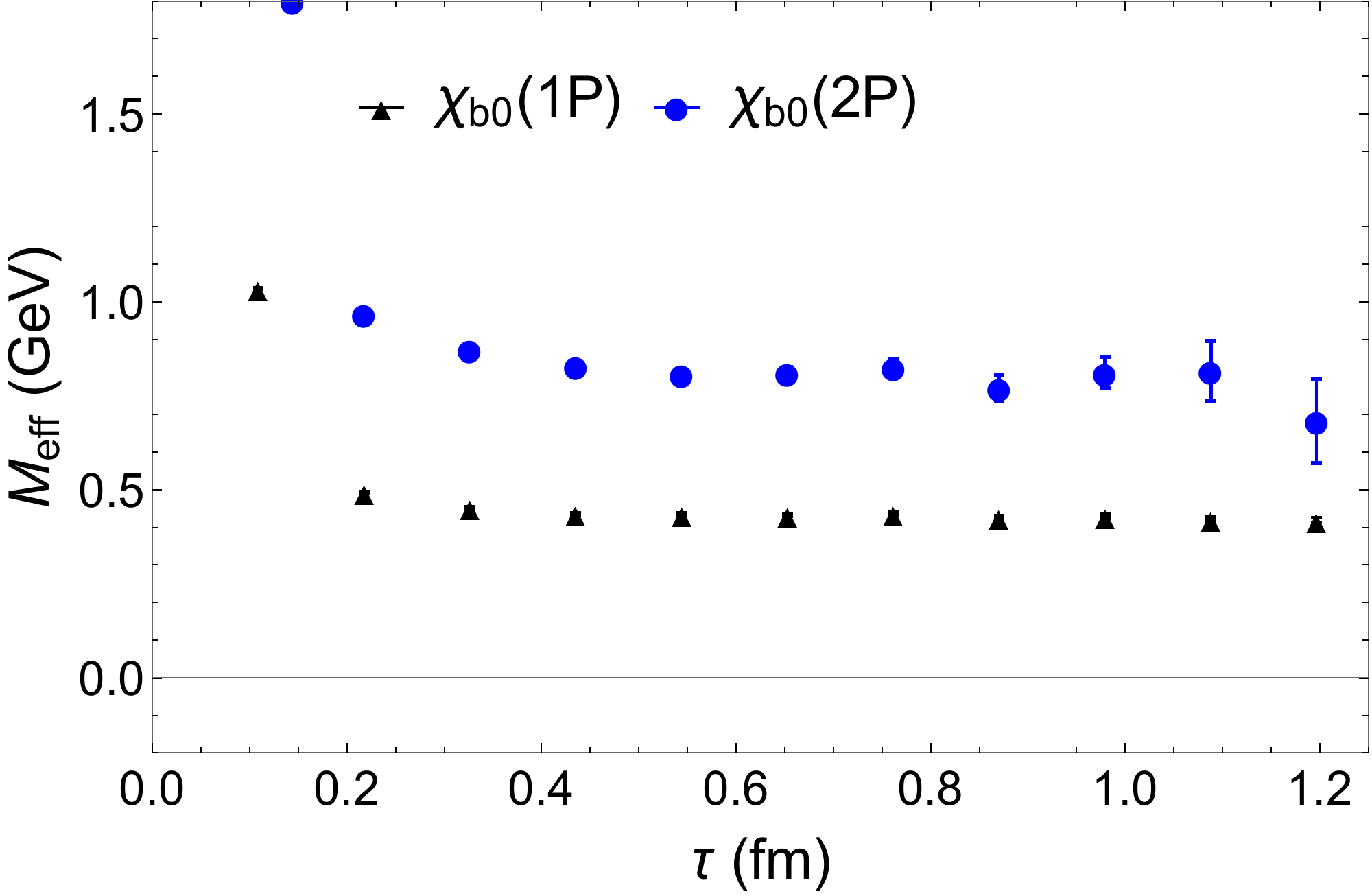}
  \caption{The effective masses, $M_\mathrm{eff}$, at $T=0$ for different $\Upsilon$ (left) and
  $\chi_{b0}$ (right) states calculated on the lattice with $a=0.109$ fm.}
  \label{fig:meff0}
\end{figure*}

In Fig. \ref{fig:meff0} we show some examples of the effective masses,
$aM_\mathrm{eff}(t)=\ln [C_{\alpha}(t)/C_{\alpha}(t+a)]$, at $T=0$. Those were found to
reach a plateau for $\tau \simeq 0.4$~fm for all states. For the excited states we
see stable plateaus up to $\tau$ of around $1.2$~fm, beyond which signal deteriorated.
Performing single-exponential fits in the plateau region we extracted the energy
levels of different bottomonium states. Since the NRQCD energy levels contain an
additional lattice spacing dependent constant, the differences of these energy levels
have the physical interpretation of mass differences of different bottomonium states.
Thus, in this work we set the spin-average energy of 1S bottomonium, $\bar
E_{1S}=(E_{\eta_b}+3 E_{\Upsilon})/4$, to be the zero of the mass/energy, and quote
masses/energies of the rest of the bottomonium states with respect to these baselines
for each lattice spacing. (Note that, in Ref.~\cite{Larsen:2019bwy} this baseline was
set by the  $\eta_b$ energy level.) The reason being $\bar E_{1S}$ remains unaffected
by the spin-spin interaction, which is difficult to reproduce accurately using
tree-level NRQCD Lagrangian used in this study. The energy differences show a very
mild dependence on the lattice spacing, and in most cases can be fitted with a
constant to obtain our final estimate for the energy differences. The only exceptions
are the energy of $\chi_{b0}$ and the 2S hyperfine splitting, where the
$a$-dependence cannot be neglected. We also fitted the resulting energy differences
with a constant plus a term proportional to $a^2$ to remove the remaining
discretization effects, this resulted in small differences in the central value but
larger statistical errors for the final estimate. We used this procedure to compare
with the experimental results, except for 3S state, where it gives too large
statistical errors. The comparison of the zero temperature results on the mass
differences with the experimental results from the Particle Data Group (PDG)~\cite{PDG18}
is shown in Table~\ref{tab:comp_pdg}. We see very good agreement between our lattice
NRQCD calculations and the experimental results within the estimated errors in most
cases. The only exception is the $\chi_{b0}(1P)$ state, where a small tension between
our result and the PDG value is observed. Using the result for the 3S hyperfine
splitting we can also predict the mass of the $\eta_b(3S)$ state to be $10341.8 \pm
6.2$~MeV, that has not yet been observed experimentally.

\begin{table}[!t]
\centering
\begin{tabular}{rrr}
  \hline
  state  & $\Delta M$~[MeV] & $ \Delta M(PDG)$~[MeV] \\ [2mm]
  $\Upsilon(3S)$   & 906.0(25.0)(5.2)   & 910.3(0.7) \\  [1mm]
  $h_b(2P)$        & 804.4(35.8)(4.7)   & 814.9(1.3) \\ [1mm]
  $\chi_{b2}(2P)$  & 809.2(36.2)(4.7)   & 823.8(0.9) \\ [1mm]
  $\chi_{b1}(2P)$  & 802.2(34.9)(4.7)    & 810.6(0.7) \\ [1mm]
  $\chi_{b0}(2P)$  & 786.8(32.7)(4.6)    & 787.6(0.8) \\ [1mm]
  $\Upsilon(2S)$   & 582.7(9.8)(3.4)    & 578.4(0.6) \\ [1mm]
  $h_b(1P)$       & 454.5(4.7)(2.6)    & 454.4(0.9) \\ [1mm]
  $\chi_{b2}(1P)$  & 463.3(4.8)(2.7)    & 467.3(0.6) \\ [1mm]
  $\chi_{b1}(1P)$  & 448.9(4.6)(2.6)    & 447.9(0.6) \\ [1mm]
  $\chi_{b0}(1P)$  & 421.3(4.7)(2.4)    & 414.5(0.7) \\ [1mm]
  hyperfine(3S)    & 13.4(6.2)(0.1)     &  NA        \\ [1mm]
  hyperfine(2S)    & 24.1(1.0)(0.1)     & 24.5(4.5)  \\ [1mm]
  \hline
 \end{tabular}
\caption{Comparisons of mass differences $\Delta M$ (in units of MeV) of various bottomonium states
with respect to the 1S spin-averaged mass obtained from our lattice calculations with
that from PDG~\cite{PDG18}. The last two rows show the 2S and 3S hyperfine
splitting. The second error in our results corresponds to the uncertainty of the
$r_1$ scale.}
\label{tab:comp_pdg}
\end{table}

\subsection{In-medium case}\label{sc:in-medium}

\begin{figure*}[!t]
  \centering
  \includegraphics[width=0.43\textwidth]{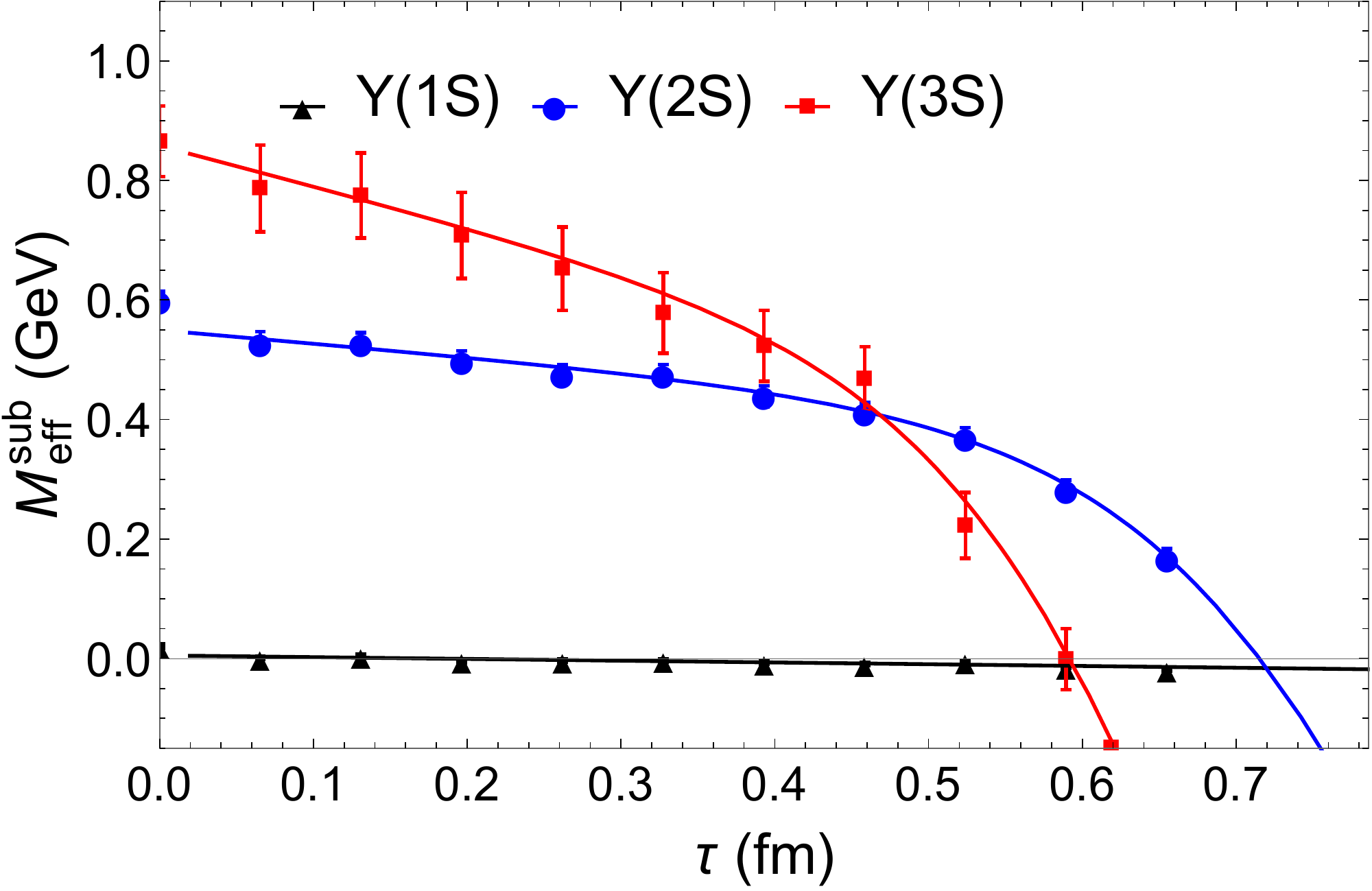}
  \hspace{0.03\textwidth}
  \includegraphics[width=0.43\textwidth]{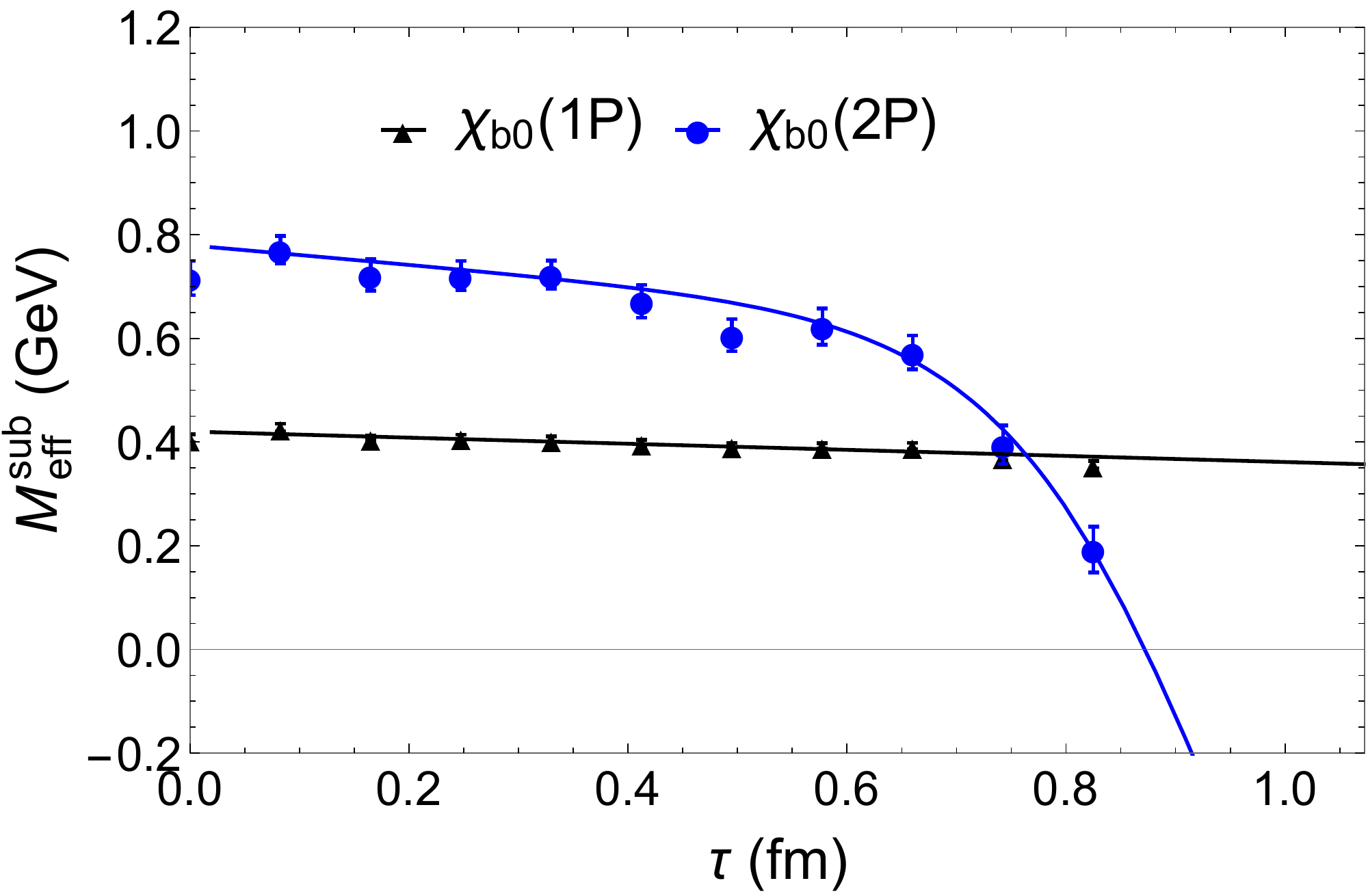}
  \caption{Continuum-subtracted effective masses, $M_\mathrm{eff}^\mathrm{sub}$, of $\Upsilon$ states at $T=251$ MeV
  (left) and  $\chi_{b0}$ states at $T=199$ MeV (right).}
  \label{fig:meff_T}
\end{figure*}

\begin{figure}[!t]
  \centering
  \includegraphics[width=0.43\textwidth]{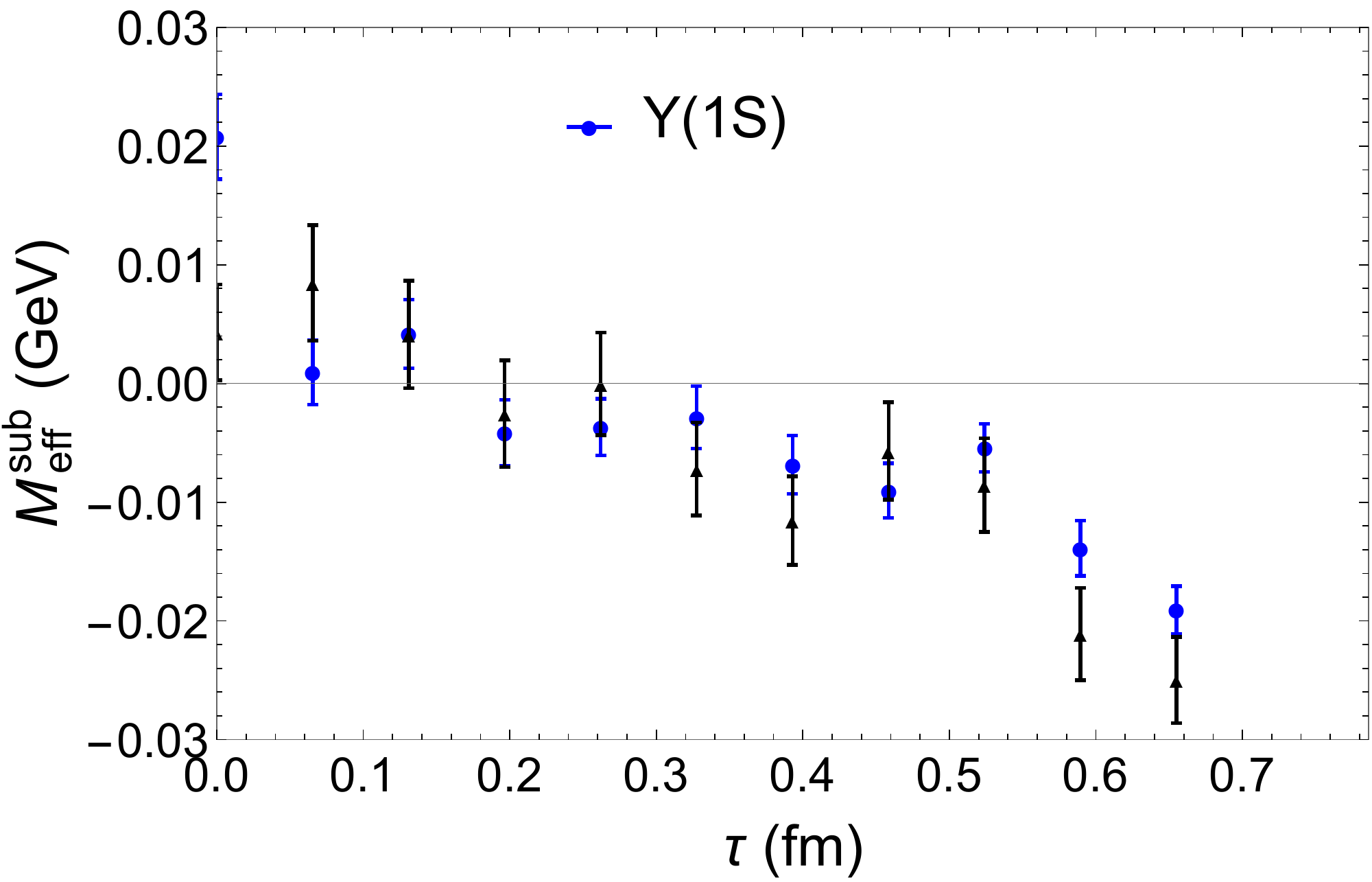}
  \caption{Comparison of the continuum-subtracted effective mass, $M_\mathrm{eff}^\mathrm{sub}$, of $\Upsilon(1S)$  at
  $T=251$ MeV obtained in this study (circle) and using Gaussian extended source
  (triangle) of Ref.~\cite{Larsen:2019bwy}.}
  \label{fig:comp2Gauss}
\end{figure}

In NRQCD the spectral function, $\rho(\omega,T)$, is related to the Euclidean time
correlation function:
\begin{equation}
  C_{\alpha}(\tau,T) = \int_{-\infty}^{\infty} d \omega \rho_{\alpha}(\omega,T)
  e^{-\omega \tau} \,.
  \label{eq:corr-sf}
\end{equation}
Here, $\alpha$ labels the bottomonium operator of interest. Bottomonium states
correspond to peaks in the spectral function having some in-medium width. At large
$\omega$ many states contribute to the spectral function, forming a continuum.
Therefore, we can write the spectral function as
\begin{equation}
  \rho_{\alpha}(\omega,T) = \rho_{\alpha}^\mathrm{med}(\omega,T) +
  \rho_{\alpha}^\mathrm{high}(\omega) \,,
\end{equation}
with the second term parameterizing the continuum part of the spectral function. In
the zero-temperature limit $\rho_{\alpha}^\mathrm{med}(\omega,T)=A_{\alpha}
\delta(\omega-M_{\alpha})$, $M_{\alpha}$ being the mass of the corresponding
bottomonium state. Here we note that the use of extended operators reduces the relative contribution of $ \rho_{\alpha}^\mathrm{high}$ \cite{Wetzorke:2001dk,Stickan:2003gh}. For this reason the effective masses in Fig. \ref{fig:meff0} approach a plateau at relatively small $\tau$. The continuum part, $\rho_{\alpha}^\mathrm{high}(\omega)$, is
expected to be temperature independent. This was seen to fit with our calculation, as the temperature
dependence of $C_{\alpha}(\tau,T)$ for $\tau \lesssim 0.3fm$~fm was very small, with the small difference being in agreement with changes due to the medium. Thus,
following Ref.~\cite{Larsen:2019bwy}, for each lattice spacing we can identify the
contribution of $\rho_{\alpha}^\mathrm{high}(\omega)$ to the correlator,
$C_{\alpha}^\mathrm{high}(\tau)$, as
\begin{equation}
 C_{\alpha}(\tau,T=0) = A_{\alpha} e^{-M_{\alpha} \tau} +
  C_{\alpha}^\mathrm{high}(\tau) \,.
\end{equation}
Here, $A_{\alpha}$ and $M_{\alpha}$ are the amplitude and mass of the corresponding
bottomonium state, and $C_{\alpha}^\mathrm{high}(\tau)$ is the Laplace transform of
$\rho_{\alpha}^\mathrm{high}$.  Using the single-exponential fits to the vacuum
correlators for $\tau\gtrsim 0.6$~fm, and subtracting off this contribution
from $C_{\alpha}(\tau,T=0)$  we isolated $C_{\alpha}^\mathrm{high}(\tau)$ for each
value of $\beta$. Further, following Ref.~\cite{Larsen:2019bwy}, for each temperature
we then defined the continuum-subtracted correlator as
\begin{equation}
  C_{\alpha}^\mathrm{sub}(\tau,T) = C_{\alpha}(\tau,T) -
   C_{\alpha}^\mathrm{high}(\tau) \,.
\end{equation}
Thus, the continuum-subtracted correlator, $C_{\alpha}^\mathrm{sub}(\tau,T)$, is
mostly sensitive to $\rho_{\alpha}^\mathrm{med}(\omega,T)$, encoding the in-medium
bottomonium properties. We then studied the in-medium bottomonium properties using
the continuum-subtracted effective masses,
\begin{equation}
 a M_\mathrm{eff}^\mathrm{sub}(\tau,T) = \ln \left(
  C_{\alpha}^\mathrm{sub}\left(\tau,T \right) /
  C_{\alpha}^\mathrm{sub}\left(\tau+a,T\right)
  \right) \,.
  \label{eq:effMsub}
\end{equation}

In Fig. \ref{fig:meff_T} we show typical examples of $M_\mathrm{eff}^\mathrm{sub}$
as a function of $\tau$--- for the $\Upsilon$ states at $T=251$~MeV and for the
$\chi_{b0}$ states  at  $T=199$~MeV. At small $\tau$,
$M_\mathrm{eff}^\mathrm{sub}$ are approximately equal to the vacuum masses. As $\tau$ increases, we
see an approximately linear decrease of $M_\mathrm{eff}^\mathrm{sub}$.
Finally, for $\tau \simeq 1/T$ we see a rapid drop-off. Similar behaviors of
$M_\mathrm{eff}^\mathrm{sub}$ for the ground states were also observed in the previous
study using Gaussian smeared meson operators~\cite{Larsen:2019bwy}. As discussed in
Ref.~\cite{Larsen:2019bwy}, the slope of the linear decrease of
$M_{eff}^\mathrm{sub}$ can be understood in terms of a thermal width. We see that the
slope is larger for higher excited  bottomonium states, i.e., the thermal width of
different bottomonium states follows the expected hierarchy of their sizes. Higher
excited states have larger size and therefore are more affected by the medium,
leading to larger width. The behavior of the effective masses at $\tau \simeq 1/T$ is
related to the tail of the spectral function at small $\omega$, and may depend on the
choice of the meson operator~\cite{Larsen:2019bwy}. Therefore, it is important to
compare the results on the subtracted effective masses obtained with different meson
operators. In Fig. \ref{fig:comp2Gauss} we compare $M_\mathrm{eff}^\mathrm{sub}$ of
subtracted $\Upsilon(1S)$ at $T=251$~MeV with  the corresponding results obtained
with Gaussian smeared sources of Ref.~\cite{Larsen:2019bwy}. Good agreement was found
between the present results and those obtained of Ref.~\cite{Larsen:2019bwy},
especially for $\tau \nsim 1/T$. Therefore, our conclusion regarding the in-medium
modification of the spectral functions is not affected by the choices of meson
operators. For $\tau \simeq 1/T$ we see a smaller drop-off in
$M_\mathrm{eff}^\mathrm{sub}$ compared to that observed in
Ref.~\cite{Larsen:2019bwy}. Thus, the small-$\omega$ tail of the spectral functions
plays a less prominent role here. We found that the behaviors of
$M_\mathrm{eff}^\mathrm{sub}$ for $\eta_b(nS)$ are very similar to that of
$\Upsilon(nS)$, and that of $\chi_{b1}(nP)$, $\chi_{b2}(nP)$ and $h_b(nP)$ are very
similar to that of $\chi_{b0}(nP)$.

As introduced in Ref.~\cite{Larsen:2019bwy}, the simplest theoretically motivated
parameterization of the in-medium spectral function that can describe the generic
behavior of $M_\mathrm{eff}^\mathrm{sub}$ observed here is as follows
\begin{align}
  \rho_\alpha^\mathrm{med}(\omega,T) = &
  A_\alpha^\mathrm{cut}(T) \, \delta \left( \omega - \omega_\alpha^\mathrm{cut}(T) \right) \nonumber \\
  + & A_\alpha(T) \exp\left( -\frac{ \left[ \omega - M_\alpha(T) \right]^2}{2 \Gamma_\alpha^2(T)} \right) \,.
\label{eq:gauss}
\end{align}
The first term in the above equation provides a simple parameterization of the
low-$\omega$ tail of the spectral function. As explained in Ref.
\cite{Larsen:2019bwy}, this tail is important for understanding the behavior of the
effective masses around $\tau \simeq 1/T$. The second term gives rise to the linear
behavior in $\tau$ of $M_\mathrm{eff}^\mathrm{sub}$, with the slope given by
$\Gamma_\alpha^2$. For each temperature, we fitted
$M_\mathrm{eff}^\mathrm{sub}(\tau)$ with the Ansatz given by Eq.~\eqref{eq:gauss},
and using Eqs.~\eqref{eq:corr-sf}, \eqref{eq:effMsub}, to determine the in-medium
masses, $M_\alpha(T)$, and width, $\Gamma_\alpha(T)$, of different bottomonium states. Since
the tail of the spectral function plays a less prominent role in the present
study,  for $T \le 173$~MeV and all temperatures for $\Upsilon$ 1S, we performed fits, setting $A_\alpha^\mathrm{cut}=0$, and
omitting 1-3 data points for the largest values of $\tau$.  Only for higher
temperatures  was the term proportional to $A_\alpha^\mathrm{cut}$ included.
 We generally find good fits with $\chi ^2 $ divided by degrees of freedom being around $0.5$. 
In some cases the data points fluctuate more than the size of the estimated errors. Examples of such cases include  $\Upsilon (1S)$ and also $\chi _{b0}$ at $199$~MeV, as can be seen on the right in Fig. \ref{fig:meff_T}. In these cases we found that $\chi ^2 $ divided by degrees of freedom was around $2$. The fit still seem to work nicely, so it is most likely the errors that were a bit too small.

\begin{figure*}[!t]
  \centering
  \includegraphics[width=0.43\textwidth]{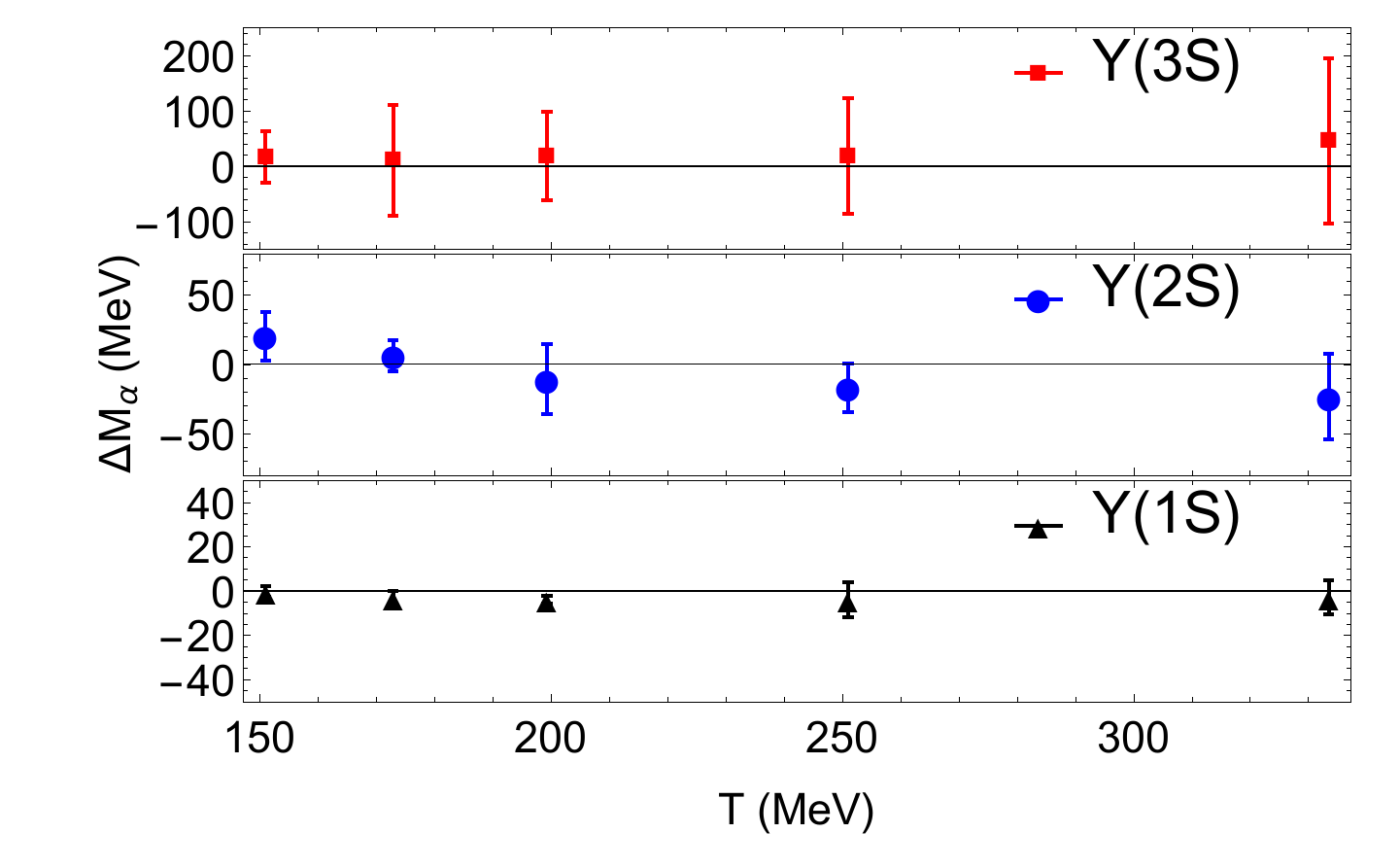}
  \hspace{0.03\textwidth}
  \includegraphics[width=0.43\textwidth]{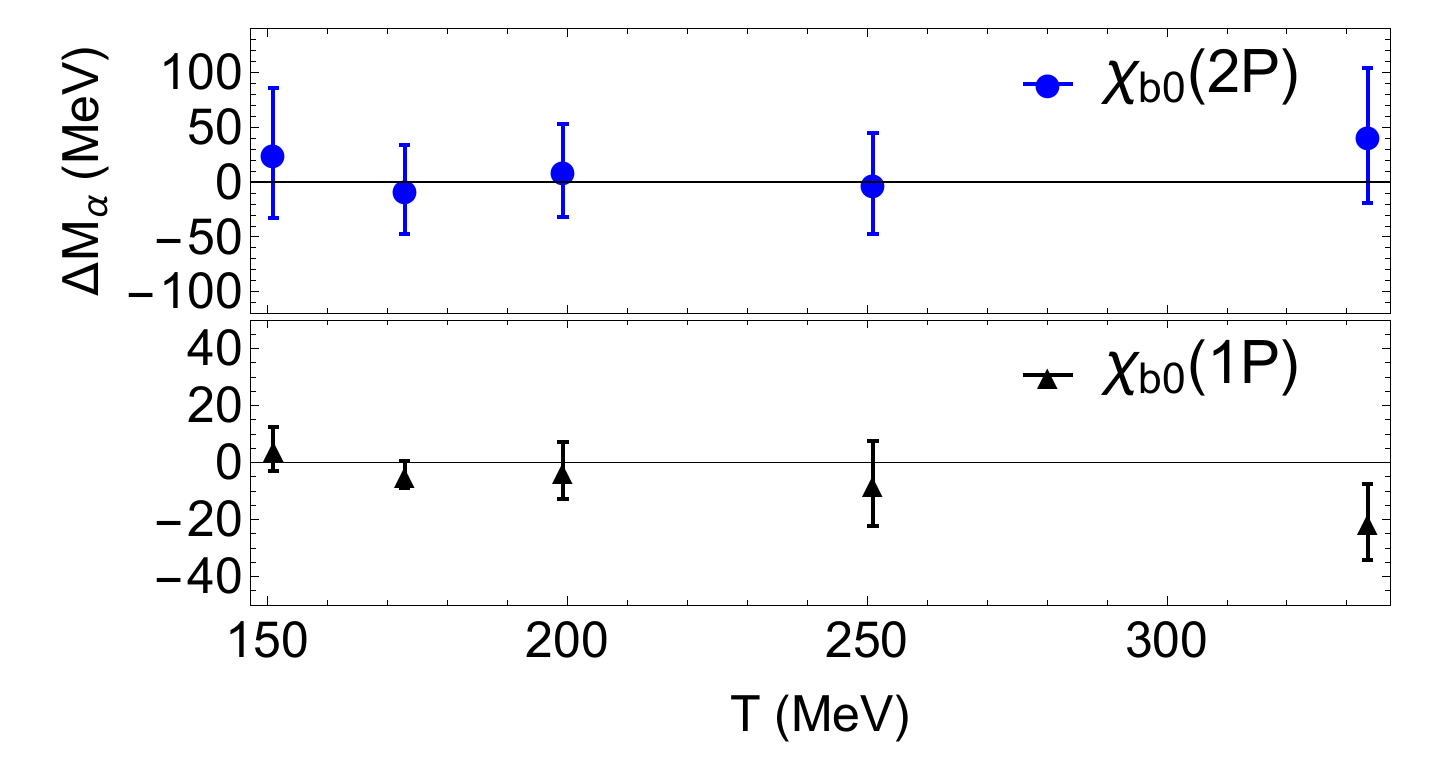}
  \caption{The change of the in-medium mass compared to the vacuum mass, $\Delta
  M_\alpha = M_\alpha(T) - M_\alpha^0$,  for the $\Upsilon$ (left) and $\chi_{b0}$
  states (right) as function of the temperature.}
  \label{fig:mass}
\end{figure*}

\begin{figure*}
  \centering
  \includegraphics[width=0.43\textwidth]{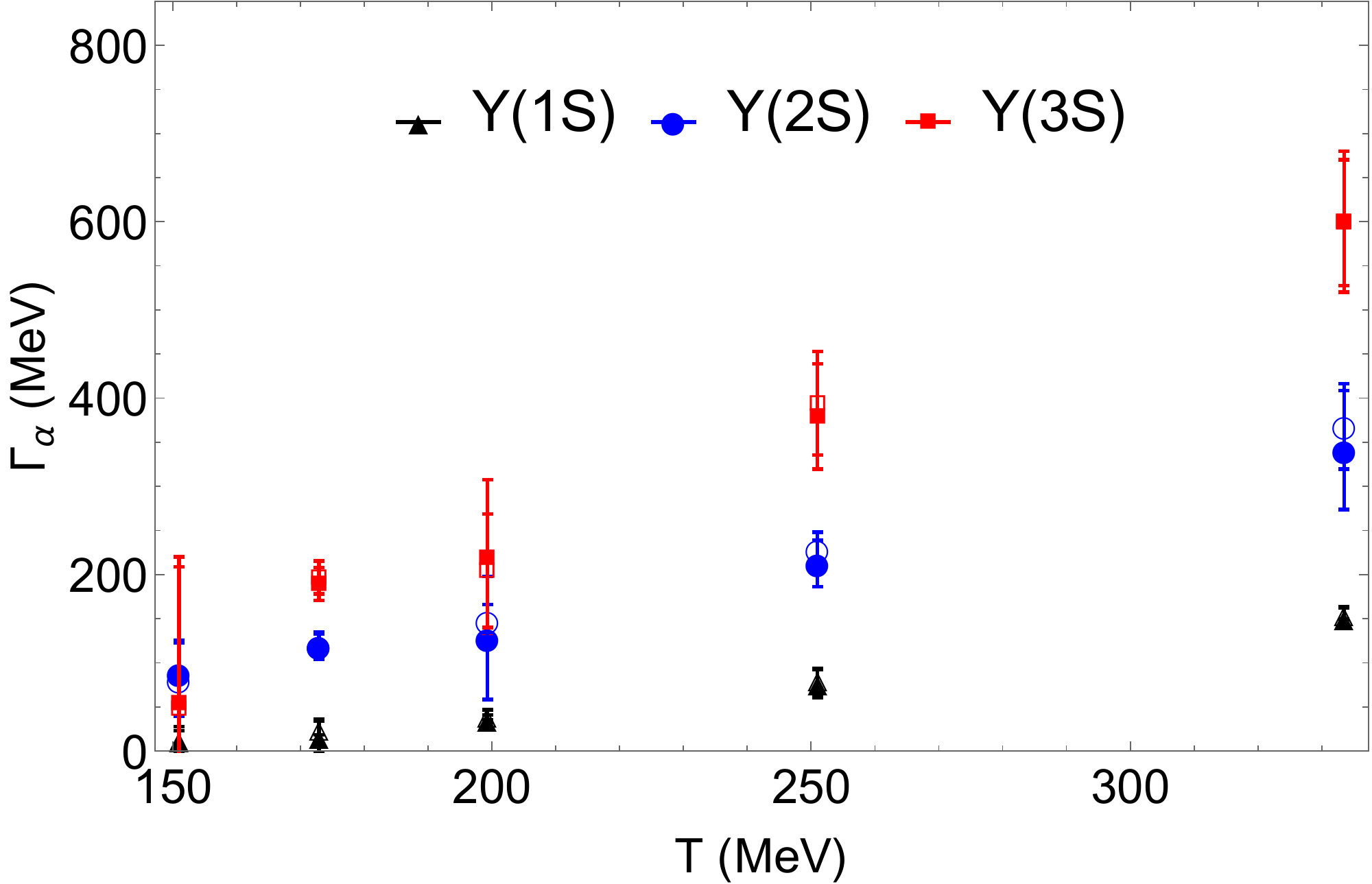}
  \hspace{0.03\textwidth}
  \includegraphics[width=0.43\textwidth]{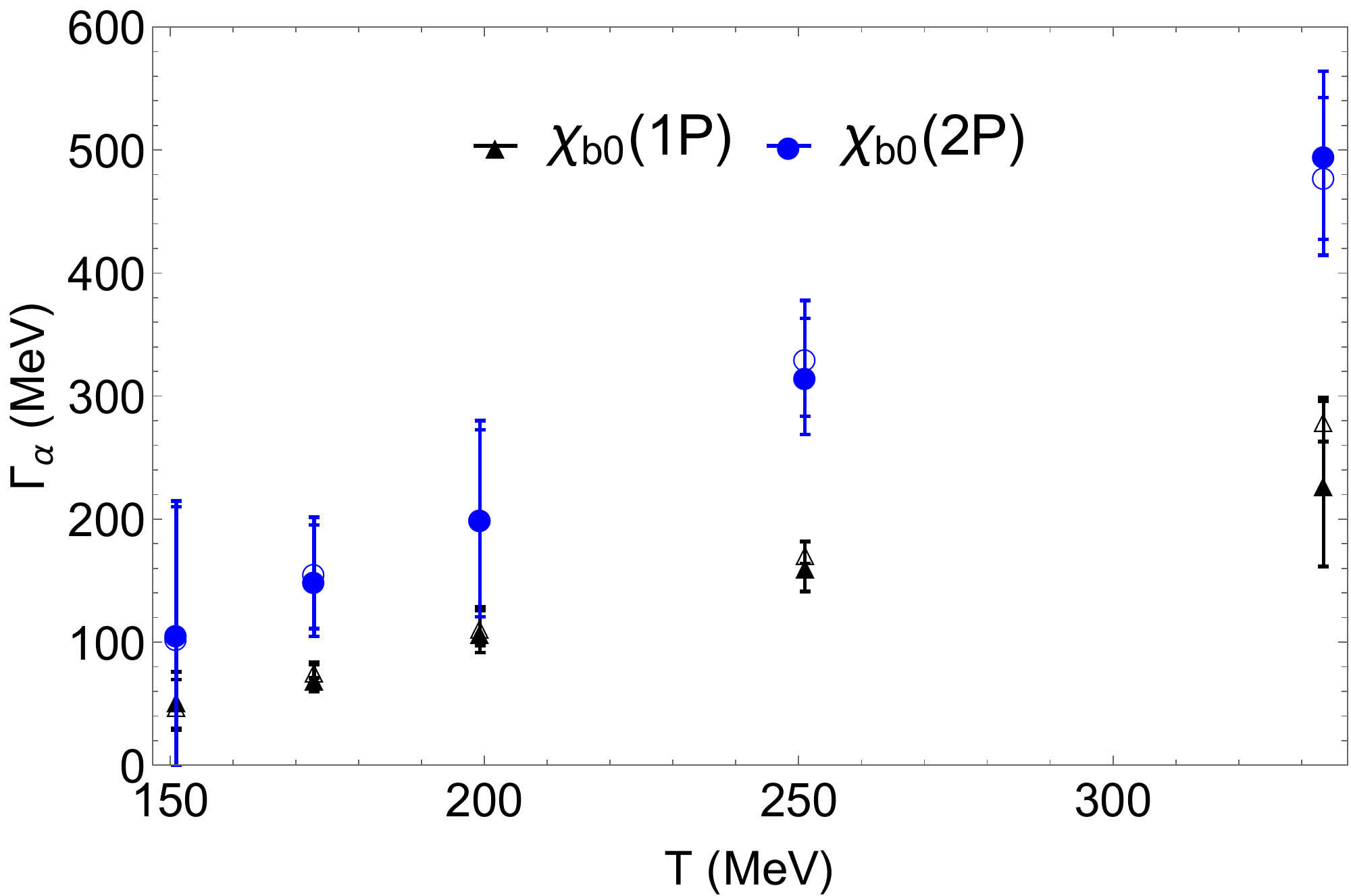}
  \caption{Thermal width, $\Gamma_\alpha$, of $\Upsilon$ (left) and $\chi_{b0}$ states
  (right) as function of the temperature using Eq. (\ref{eq:gauss}) (filled) and using Eq. (\ref{eq:3peaks}) (empty). Fit with Eq. (\ref{eq:3peaks}) assumes  $M_\alpha(T) = M_\alpha^0$ and error bars are thus slightly smaller.}
\label{fig:width}
\end{figure*}

The change of the in-medium mass parameter compared to the vacuum mass
($M_\alpha^0$), $\Delta M_\alpha(T) = M_\alpha(T) - M_\alpha^0$, and width parameter,
$\Gamma_\alpha(T)$, are shown in Figs. \ref{fig:mass} and \ref{fig:width},
respectively. The in-medium masses of different states obtained from the fits turned
out to be very similar to the vacuum masses. In fact, we do not see any statistically
significant deviations from the $T=0$ results. On the other hand, $\Gamma_\alpha(T)$ shows a
clear increase with increasing temperature. For large enough temperatures, $\Gamma_\alpha(T)$
appears to approximately rise linearly with $T$. Nearly for the entire $T$-range, the
in-medium width was found to follow the sequential hierarchical pattern according to
the increasing sizes of the bottomonium states:  $\Gamma_{1S}(T) < \Gamma_{1P}(T) <
\Gamma_{2S}(T) < \Gamma_{2P}(T) < \Gamma_{3S}(T)$. Moreover, $\Gamma_{3S}\gtrsim
M_{3S}-M_{2S}$ and $\Gamma_{2P}\gtrsim M_{2P}-M_{1P}$ for $T\gtrsim200$~MeV. As a
result, at these temperatures $2S$ and $3S$, as well as the $1P$ and $2P$ states
will together appear as  broad structures in their respective spectral functions.
These observations lead us to conclude that, similar to what have been observed in
the experiments~\cite{Chatrchyan:2011pe,Chatrchyan:2012lxa}, for $T\gtrsim200$~MeV it
will become difficult to individually identify the $2S$, $3S$, $1P$ and $2P$ states
within the experimentally measured line shapes of the invariant-mass distributions.

Lastly, we address the question to what extent the estimated thermal widths of
bottomonium states depend on the model for the spectral function used to interpret
our lattice QCD results presented here. Since $M_\mathrm{eff}^\mathrm{sub}(\tau)$
show a linear in $\tau$ behavior and we do not observe any significant thermal mass
shift, following Ref.~\cite{Larsen:2019bwy}, we also used the following model for the
spectral function:
\begin{align}
  \rho_\alpha^\mathrm{med}(\omega,T) = &
  A_\alpha^\mathrm{cut}(T) \, \delta( \omega - \omega_\alpha^\mathrm{cut}(T) ) \nonumber \\
  + & \delta( \omega - M_\alpha^0 + \Delta_\alpha(T) ) \nonumber \\
  + & \delta( \omega - M_\alpha^0 ) \nonumber \\
  + &  \delta( \omega - M_\alpha^0 - \Delta_\alpha(T) ) \,.
  \label{eq:3peaks}
\end{align}
Here, $M_\alpha^0$ is the vacuum bottomonium mass, and the parameters
$A_\alpha^\mathrm{cut}$ and $\omega_\alpha^\mathrm{cut}$ describe the low-$\omega$
tail of the spectral function. For $T\le 173$ MeV again we used
$A_\alpha^\mathrm{cut}=0$. The equivalent thermal width in this case is $\Gamma_\alpha(T) =
\sqrt{2/3} \Delta_\alpha(T)$. Carrying out fits with the above Ansatz we obtained
thermal widths that, as shown in Fig. \ref{fig:width}, within errors, agreed with the ones obtained by using the
Gaussian Ansatz. Therefore, our estimates of thermal width do not depend very much on
the precise functional form of the fit Ansatz.

\section{Conclusion}\label{sc:conclusions}

For the very first time, we studied in-medium properties up to $3S$ and $2P$ excited
bottomonium states using lattice QCD at temperatures $T\simeq 150-350$~MeV. This
lattice QCD study was made possible through the introduction of novel bottomonium
operators within the lattice NRQCD framework, and implementation of a variational
analysis based on these novel operators. We found that the effective masses
constructed out of the continuum-subtracted bottomonium correlation functions drop
off linearly in Euclidean time. We argued that the behaviors of the
continuum-subtracted effective masses can be understood in terms of a couple of
theoretically-motivated, simple models of the bottomonium spectral functions.
For all of the models considered, we found indications of thermal broadening of bottomonium
states in QGP.  For the entire temperature range, the magnitudes of the thermal
broadening were found to follow the expected sequential hierarchical pattern
according to the increasing sizes of the bottomonium states. Further, we found that
for $T\gtrsim 200$~MeV the thermal broadening of the $2S$, $3S$, $1P$ and $2P$ states
becomes large enough that it would be difficult to identify these states separately
within the corresponding spectral functions.

\vspace{-0.4em}
\section*{Acknowledgments}
\vspace{-0.2em}

This material is based upon work supported by the U.S. Department of Energy, Office
of Science, Office of Nuclear Physics: (i) Through the Contract No. DE-SC0012704;
(ii) Through the Scientific Discovery through Advance Computing (SciDAC) award
Computing the Properties of Matter with Leadership Computing Resources. (iii) Stefan
Meinel acknowledges support by the U.S. Department of Energy, Office of Science,
Office of High Energy Physics under Award Number DE-SC0009913.

  This research used awards of computer time: (i) Provided by the USQCD consortium at
  its Fermi National Laboratory, Brookhaven National Laboratory and Jefferson
  Laboratory computing facilities; (ii) Provided by the INCITE program at Argonne
  Leadership Computing Facility, a U.S. Department of Energy Office of Science User
  Facility operated under Contract No. DE-AC02-06CH11357;  (ii) Provided by the ALCC
  program at National Energy Research Scientific Computing Center, a U.S. Department
  of Energy Office of Science User Facility operated under Contract No.
  DE-AC02-05CH11231;  (iii) Provided by the INCITE programs at Oak Ridge Leadership
  Computing Facility, a DOE Office of Science User Facility operated under Contract
  No. DE-AC05-00OR22725.

\bibliographystyle{elsarticle-num}

\end{document}